\documentstyle[aps,twocolumn,epsfig]{revtex}
\newcommand{\beq}{\begin{equation}}
\newcommand{\eeq}{\end{equation}}
\newcommand{\bea}{\begin{eqnarray}}
\newcommand{\eea}{\end{eqnarray}}
\newcommand{\lsimeq}{\stackrel{<}{\scriptstyle\sim}}
\newcommand{\gsimeq}{\stackrel{>}{\scriptstyle\sim}}

\newcommand{\bml}{\begin{mathletters}}
\newcommand{\eml}{\end{mathletters}}
\newcommand{\commentout}[1]{{}}
\newcommand{\k}{{\bf k}}

\newcommand{\p}{{\bf p}}

\newcommand{\etal}{{\it et al.}}
\newcommand{\eq}[1]{(\ref{#1})}

\newcommand{\rb}{$^{85}$Rb }

\begin{document}
\flushbottom
\wideabs
{
\title{Rapid Adiabatic Passage from an Atomic to a
Molecular Condensate
}
\author{Matt Mackie,$^1$ Andrew Carmichael,$^3$
Marijan Ko\u{s}trun,$^3$ Rory J. Perkins,$^3$ Chen
Xu,$^3$ Yi Zhen,$^3$ \\
Kalle-Antti Suominen,$^{1,2}$ and Juha
Javanainen$^{3,4}$}
\address{
$^1$Helsinki Institute of Physics, PL 64, FIN-00014
Helsingin yliopisto, Finland \\
$^2$Department of Physics, University of
Turku, FIN-20014, Turun yliopisto, Finland \\
$^3$Department of Physics, University of Connecticut,
Storrs, Connecticut, 06269-3046\\
$^4$Optics and Molecular Materials, PL 2200,
FIN-02015 Helsinki University of Technology, Finland
}
\date{\today}
\maketitle

\begin{abstract}
We examine collective magnetoassociation of a
Bose-Einstein condensate (BEC),
focusing on rapid adiabatic passage from atoms to
molecules induced by
a sweep of the magnetic field across a
wide ($\gsimeq
10\text{ G}$) Feshbach resonance in ${}^{85}$Rb. This
problem raises
an interest because
strong magnetoassociation is expected to favor the
creation of
molecular-dissociated atom pairs over the formation of
molecular
BEC {[Javanainen and Mackie, \prl {\bf 88}, 090403
(2002)]}.
Nevertheless, the conversion to atom pairs is
found to depend on
the direction of the sweep, so that a system initially
above threshold
(open dissociation channel) may in fact give
efficient conversion to
molecules.
\end{abstract}
\pacs{PACS numbers: 03.75.Fi,03.65.Bz}
}

Photoassociation occurs when a laser is on resonance
with a transition from the collisional state of an
atom pair to the
bound state of a molecule. The analogous
process of
magnetoassociation, which involves a Zeeman-shifted
molecular level
on Feshbach resonance with a colliding atom
pair, has an
identical formalism, and intuition developed in one
case is applicable
to the other. Specifically, the probability of
photoassociation in a
Bose-Einstein condensate (BEC) was found to approach
unity~\cite{UNIT_EFF}, which opened the door to the
possibility of using
coherent magnetoassociation~\cite{MA_MBEC} or
photoassociation~\cite{JAV99,RAMAN} of
already-Bose-condensed atoms to
create a molecular BEC.

Early theory of coherent association was implicitly
based on
few-mode models that neglect
transitions to noncondensate
modes~\cite{MA_MBEC,JAV99,RAMAN}. Such
``rogue"~\cite{KOS00,JAV02}, or
unwanted~\cite{BOH99,OTHER_ROGUES},
transitions to noncondensate modes occur
because the dissociation of a zero-momentum ($\k=0$)
BEC molecule need
not take the atoms back to the zero-momentum atomic
condensate,
but may just as well
end up with two atoms with opposite momenta
($\pm\k$). Since the coherent
condensate coupling scales like the square root of
the laser intensity
(Feshbach-resonance width) and the noncondensate
coupling scales
like the intensity, rogue
dissociation is
expected to play a dominant role in strongly coupled
atom-molecule systems~\cite{KOS00,JAV02}.

Pioneering experiments~\cite{WYN00} with
photoassociation of $^{87}$Rb BEC were found to be
just on the
verge~\cite{KOS00} of coherent atom-molecule
conversion. Next-generation
Na~\cite{MCK02} and $^7$Li~\cite{PIC02} experiments
were aimed
at the strongly interacting regime, and probed the
potential
photoassociation rate limit. Meanwhile,
experiments on magnetoassociation in
\rb have led to dramatic losses of BEC atoms for a
sweep of the magnetic
field across resonance~\cite{COR00}, a collapsing
condensate with bursts
of atoms emanating from a remnant
condensate~\cite{DON01}, a
counterintuitive {\em decrease} in condensate losses
for an increasing
interaction time~\cite{CLA02}, and collective
burst-remnant
oscillations~\cite{DON02}.

Rogue dissociation is the key to an overall
understanding of
these experimental results. First, although the
experiments remain inconclusive on this
score~\cite{MCK02,PIC02}, a rate limit is to be
expected when
the conversion to rogue-dissociated atom pairs
dominates over
the formation of molecular
condensate~\cite{JAV02,BOH99}.
Similarly, the losses
for across-resonance sweeps~\cite{COR00}, and the
counterintuitive
losses~\cite{CLA02}, can be viewed as rapid
adiabatic passage from BEC to rogue-dissociated atoms
pairs~\cite{MAC02}.
The remnant-burst oscillations~\cite{DON02} are
identified as Ramsey
fringes in the evolution between an atomic condensate
and a molecular
condensate dressed by noncondensate atom
pairs~\cite{MAC02,KOK02,KOE02,DUI02}.

There remains the matter of the formation of a
molecular
condensate. In magnetoassociation experiments, the
sum of
the remnant and  burst atoms does not entirely account
for the initial
condensate, and the missing atoms
are roughly consistent with
calculations~\cite{MAC02,KOK02,KOE02,DUI02} of the
number of molecules formed, thus
indirectly confirming
the realization of molecular BEC.
Either way,
the (measured) calculated fraction of (atom loss)
molecular condensate is small ($\,\lsimeq$~10\%).

The question is whether an improvement of the
$^{85}\rm{Rb}_2$ conversion efficiency is possible.
Given the
existing results
\cite{JAV99,KOS00,JAV02,BOH99,OTHER_ROGUES,PIC02,DON01,CLA02,DON02,MAC02},
the immediate answer is no. The 155~G
magnetoassociation resonance in
${}^{85}$Rb is exceptionally strong, and rogue
dissociation is apparently prominent. Indeed, the
qualitative agreement
between experiment~\cite{COR00} and
theory~\cite{MAC02}
indicates that the obvious solution, a
$\sim$~1~ms sweep of the
magnetic field across resonance, merely causes
rapid adiabatic passage
from the initial atomic condensate to
rogue-dissociated atoms,
with very little molecular BEC formed.
The present Letter reports our startling observation
that,
simply by reverting the direction of the sweep of the
magnetic field from the experiments of
Ref.~\cite{COR00},
rapid adiabatic passage~\cite{JAV99,KOS00,LZMA} may
nonetheless efficiently convert from an atomic
to a molecular condensate.

We continue to refine our mean-field model for
photoassociation and Feshbach
resonance~\cite{JAV02,MAC02} in an attempt to extract as
much quantitative accuracy out of it as possible.
Since Bose enhancement favors the return of
dissociated atoms back to the molecular BEC, we
ignore all noncondensate modes of the molecule. The
model thus includes an atomic condensate [described by
the complex amplitude
$\alpha$], a molecular condensate [$\beta$], and
noncondensate atom
pairs with zero total momentum [$C(\epsilon)\propto
\langle a_\p
a_{-\p} \rangle$, $\epsilon = \hbar\k^2/m$].  As new features we now 
incorporate
possible Bose enhancement in the conversion of molecules
to noncondensate atom pairs, conversion of the
``anomalous'' atom pair amplitudes [$\langle a_{\bf
p}a_{-\bf p}\rangle$] to usual atomic probabilities
[$P(\epsilon) = \langle a^\dagger_\p a_\p\rangle$], and
a gradual high-energy cutoff of the coupling of
molecules to atom pairs [$f(\epsilon)$]. The system to
be solved reads
\bml
\bea
i\dot\alpha &=&
-{\Omega\over\sqrt{2}}\,\beta\alpha^*,\label{EQ1}\\
i\dot\beta&=&\delta_0\beta-{\Omega\over\sqrt{2}}\,\alpha\alpha
\nonumber\\
&&-{\xi\over \sqrt{2\pi}}\int_0^\infty
d\epsilon\,\sqrt[4]{\epsilon}
\sqrt{2P(\epsilon)+1}\,f(\epsilon) C(\epsilon),
\label{EQ2}
\\
i\dot C(\epsilon)&=&\epsilon\,C(\epsilon)
-{\xi\over\sqrt{2\pi}}\sqrt[4]{\epsilon}
\sqrt{2P(\epsilon)+1}f^*(\epsilon)\beta,
\label{EQ3}\\
i\dot P(\epsilon)&=&
{2(2\pi)^{3/2}\sqrt{2P(\epsilon)+1}\over\sqrt[4]{\epsilon}}\,
\Im[\xi f(\epsilon)C(\varepsilon)\beta^*]\,.
\label{EQ4}
\eea
\label{EQM}
\eml
The ``bare" energy of
the bound
molecular state referenced to the dissociation
threshold is $\hbar\delta_0$, the coupling between atomic and
molecular condensates is $\Omega$, and the coupling for
dissociation of molecules into noncondensate atom
pairs is $\xi$. The couplings
are related in a density-dependent manner:
$\xi = \Omega/(2\sqrt\pi\omega_\rho^{3/4})$,
$\omega_\rho = \hbar\rho^{2/3}/m$. The atom pair
amplitude
$C(\epsilon)$ is scaled so that Eqs.~\eq{EQM} preserve
the
norm $|\alpha|^2+|\beta|^2+\int
d\epsilon\,|C(\epsilon)|^2=1$. Also new is an average
over the density profile of trapped atoms, which
is identical for all three-dimensional Gaussian
distributions
with the same peak density~$\rho_0$.

The energy profile $f(\epsilon)$ is the most
significant new
addition.  The default shape
is such that $f(\epsilon)$ starts as $f=1$ at
$\epsilon=0$, and falls off smoothly with energy over
a characteristic scale $\hbar\,\Delta\epsilon$.
Physically, this high-energy (-momentum)
cutoff accounts for the fact that the atom-molecule
coupling is not a contact interaction, but has a
nonzero range. We use simple Fourier transform
methods~\cite{MAC02} to study the coupling-induced
renormalization of the detuning,
$\delta_0\rightarrow\delta$,
and the binding energy $E_B$ of the dressed molecules
consisting
of a coherent superposition of the bare molecules
[$\beta$] and
correlated atom pairs [$C(\epsilon)$]. The physical
detuning varies
with the magnetic
field applied on the sample as $\hbar\delta =
{(B_0-B)\Delta\mu}$, where
$B_0 = 154.9\,{\rm G}$ is the position of a particular
Feshbach resonance
in ${}^{85}$Rb, and $\Delta\mu =
2.23\,\mu_B$~\cite{KOK02} is the
difference in the magnetic moments of the participant atomic and
molecular states.

To compare with close-coupled
calculations\cite{DON02,KOK02}, we have
tried a number of different coupling
functions
$f(\epsilon)$, both ad-hoc and forms originating from
square-well and
Lennard-Jones models. For the generic $f(\epsilon)$, there is
little qualitative difference between various choices. So far, the best
match is for the ad-hoc profile
$|f(\epsilon)|^2 =
\theta(\epsilon_M-\epsilon)\,\Delta\epsilon/(\Delta\epsilon+\epsilon)$.
The abrupt cutoff at
$\epsilon_M = 20\,(\mu{\rm s})^{-1}$ is for
numerical purposes only, and was set high enough so that
it has a minor effect on the results. The remaining
parameter $\xi$ and $\Delta\epsilon$ are chosen in such
a way that the desired binding energy is
obtained at two widely spaced magnetic fields. The
parameter values used here are $\xi =
19.32\,(\mu{\rm s})^{-1/4}$ and
$\Delta\epsilon=0.9537\,(\mu{\rm s})^{-1}$.
Comparison with Ref.\cite{KOK02} shows a maximum
mismatch of
about 9\% around
160 G, which occurs because the close-coupling
calculations were matched
at the end points of the magnetic-field range
considered
($\sim$~157-162~G).

\begin{figure}
\centering
\epsfig{file=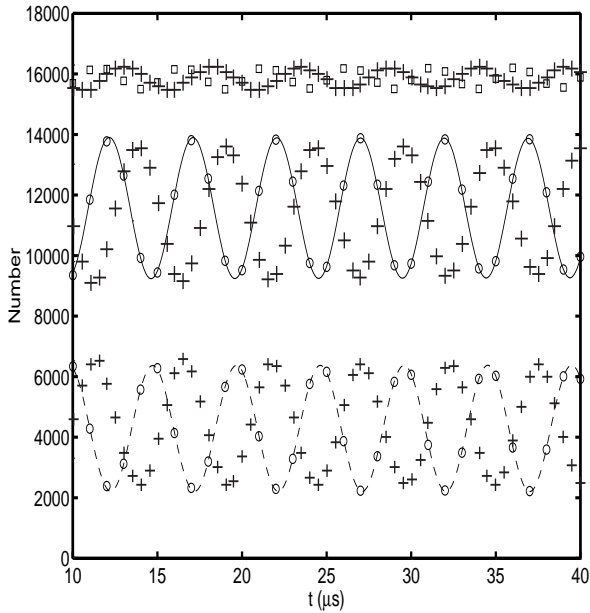,width=8.5cm,height=8.5cm}
\caption{Ramsey fringes in atom-molecule conversion,
as calculated herein (+) and in
Ref.~\protect\cite{KOK02}
($\diamond$,--,--$\,$--,$\Box$).
 From the bottom up, we have the number of
noncondensate
and condensate atoms, and the sum thereof. The Ramsey
pulses are as
in Fig.~2 of Ref.~\protect\cite{KOK02}, except that
the hold field
was 159.84~G. The peak density is
$\rho_0=1.1\times10^{13}\,{\rm cm}^{-3}$.}
\label{F2}
\end{figure}

In Ref.~\cite{KOK02}, an explicit sequence of two
magnetic field pulses is specified with the
implication
that it was used to simulate a Ramsey-fringe
experiment~\cite{DON02}. The fraction of condensate
atoms,
noncondensate atoms, and the sum thereof at
the end of the pulse sequence are given as a function
of the
time that the magnetic field is held at a constant
value
between the pulses. However, the results
in Ref.~\cite{KOK02} are not consistent with
the magnetic field during the holding period displayed
in the
pulse sequence, 160~G, and we replace this by the
value
used in the experiment being simulated, 159.84~G. We have thus carried out
the same calculations using the
present methods. In Fig.~\ref{F2} we have copied the
results
from Ref.~\cite{KOK02}, and overlaid our calculations
as plus
signs. Though the comparison may look
alarming, the apparent discrepancy originates from an
about
10\% difference in the frequency of the Ramsey
fringes, exactly as expected based on the
binding energy difference discussed above. Given that
the two
approaches are seemingly quite different, the
match of the Ramsey fringes per se is excellent.

\begin{figure}
\centering
\epsfig{file=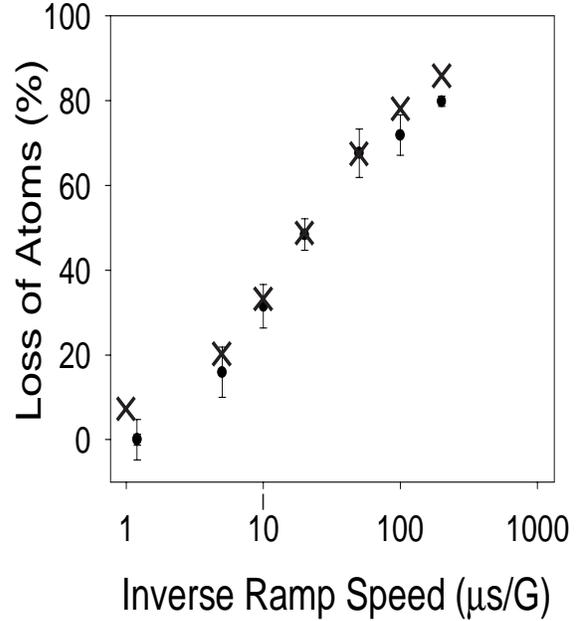,width=8.5cm,height=8.5cm}
\caption{Experimental ($\bullet$, error
bars)~\protect\cite{COR00} and
theoretical ($\times$) fraction of atoms remaining
after the magnetic
field is swept linearly from 132~G to 162~G, versus
the inverse sweep
rate. The theoretical (experimental) peak density of
the condensate is
$\rho_0=1.1\times10^{13}\,{\rm cm}^{-3}$
($\rho_0=1.0\times10^{12}\,{\rm cm}^{-3}$).
}
\label{F3}
\end{figure}

Unfortunately, when we turn to quantitative
comparisons with experimental
results, we still~\cite{MAC02} receive a mixed
message.
For instance, let us
revisit the experiment of Ref.~\cite{COR00}. The
magnetic field is swept
from 162~G to 132~G, so that the molecular dissociation channel is
initially closed and opens when resonance is
crossed. In Fig.~\ref{F3}, we borrow Fig.~2
from Ref.~\cite{COR00} showing the experimental
fraction
of atoms remaining
in the atomic condensate after the magnetic field
sweep as a function of
the inverse of the sweep rate, and overlaid our
results as crosses. The
agreement would be excellent, had we not used
the
peak density
$\rho_0=1.1\times10^{13}\,{\rm cm}^{-3}$ (the same
as in Fig.~\ref{F2}) whereas Ref.\cite{COR00} reports
$\rho_0=1.0\times10^{12}\,{\rm cm}^{-3}$.

Let us continue just the same to use the peak density
$\rho_0=1.1\times10^{13}\,{\rm cm}^{-3}$, and ask what
happens if
the magnetic field is swept in the opposite direction,
e.g.,
$142 \rightarrow 172$~G. At first blush, not much;
overlaying the
lost-atom fraction on Fig.~\ref{F3} would give
agreement with the crosses
for all but the fastest-sweep data point.

\begin{figure}
\centering
\epsfig{file=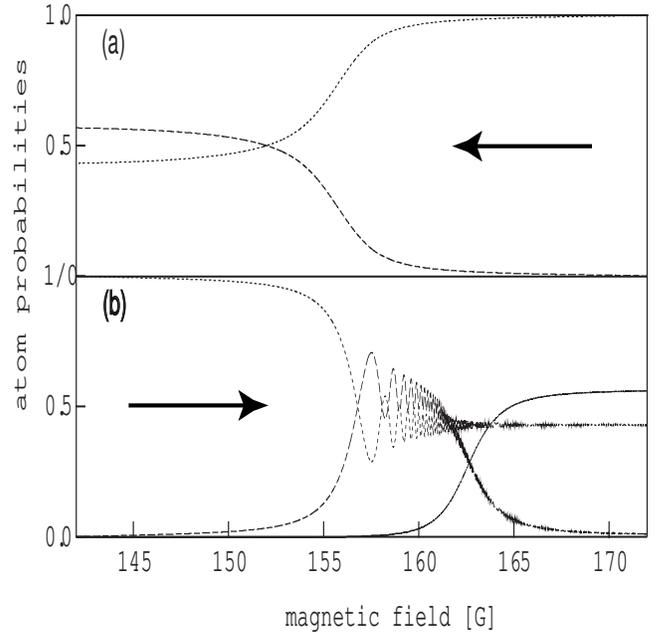,width=8.5cm,height=8.5cm}
\caption{Fraction of atoms in atomic condensate
(dotted line), in noncondensate atoms (dashed line),
and in the
molecular condensate (solid line) present while the
magnetic field is
swept by 1~G per 20~$\mu$s in the direction indicated
by the arrow in
each pane. Molecules are too few to resolve in the
upper pane. There is
no density average; the fixed density
$\rho=3.9\times10^{12}\,{\rm
cm}^{-3}$ is what the average density would be for the
peak density $\rho_0=1.1\times10^{13}\,{\rm
cm}^{-3}$.}
\label{F4}
\end{figure}

However, a profound difference emerges when the
progress of sweeps is followed in time. In
Fig.~\ref{F4} we
show the fractions of atomic and molecular condensate,
and noncondensate
atoms, as a function of the magnetic field. Pane (a)
is for a $172 \rightarrow 142$~G sweep, pane (b) for
the opposite sweep,
and both inverse
sweep rates are 20~$\mu$s/G.
In a Cornish-type~\cite{COR00} sweep, condensate atoms
are unceremoniously
converted to rogue atoms, and the molecular
fraction remains
negligible~\cite{MAC02}. An
opposite sweep begins the same; but, as the system
moves well
past the resonance, nearly all noncondensate atom
pairs are converted into molecules. This conclusion is
independent
of sweep rate, and survives the density averaging. For
the
slowest sweep rate, 85\% of the atoms would have been
converted to
molecules and 1\% to noncondensate atoms.

To intuitively understand the double-adiabatic process
that takes
place here, consider a two-level system comprised
of an atomic BEC (state 1) and  a molecular BEC
dressed by
noncondensate atom pairs (state 2). Near resonance state
2 is
primarily atom pairs, and so an adiabatic
sweep of the detuning
into the neighborhood of the threshold will necessarily
convert the initial BEC
into noncondensate atom pairs, regardless of direction
of the sweep. The asymmetry arises because far
below threshold state 2 is primarily molecular BEC. A
sweep from above to below threshold will
eventually convert the noncondensate pairs into
molecular condensate. There is no such
possibility for a sweep going the opposite way.

More precisely, our hypothesis for 
a sweep of the detuning from an initially-open to a
finally-closed
dissociation channel is as follows.
The molecule may
be strongly mixed with the atom pairs; but, the
zero-energy atomic condensate holds the initial
probability, and
is off resonance from the dressed eigenstates with any
substantial molecular component. As the detuning is
swept,
eventually eigenstates with a significant molecular
amplitude are
sufficiently close to resonance
with the atomic condensate for transitions to take
place. Transitions go mainly to the dressed molecule,
{\em the} negative-energy eigenstate
of the system. Continuing the sweep then
gives adiabatic conversion from dressed to bare
molecules.
When the detuning
is swept in the opposite direction (say, $162
\rightarrow
132$~G), so that a
dissociation channel opens for molecules after the
resonance, the
initial events are similar as in a sweep going
the other way.
However, after the resonance is crossed, the bare
molecules will not
separate as an eigenstate of their own but remain
diluted by the atom
pairs.

Whereas detuning-sweep conversion in a two-mode model is
independent of sweep direction, strong rogue
dissociation implies that transfer to molecules only
happens for a sweep in which the molecular dissociation
channel closes after the resonance. Past rate theory
reasonably surmises that, after
the dissociation channel closes, the molecules should
not dissociate
anymore~\cite{JAV99}; however, while this
serves as a convenient mnemonic, our present
result is about {\em coherent evolution}, and at least
directly is not about rates or a closing dissociation
channel.

While excellent quantitative agreement with
the Ref.~\cite{KOK02} calculations is found, we are experiencing
persistent
problems when trying to match
experiments~\cite{COR00,CLA02}. Missing
are both quenching of the molecules and collisional
decoherence of
atom pairs. No quantitative data on such
processes are available at this time, but we have been
experimenting by adding relaxation mechanisms into
our
calculations. With a hand-picked model for
collisions, an adequate
agreement with the experimental data in Fig.~\ref{F3} is reached
without adjusting the density in the calculations. The
problem is that a
universal model for collisions that would lead to
an agreement between theory and all of the
experimental data at once is elusive.

Nevertheless, we propose making a $^{85}\rm{Rb}_2$
condensate simply by
asymmetric adaptation of the single-pulse experiments
of Claussen et al.~\cite{CLA02}: form a BEC at 166~G;
sweep fast
($\sim$~1 G/$\mu$s) to 142~G, losing a negligible
number of atoms
($\sim$ 1~\%);
sweep slowly ($\sim5\times10^{-3}$~G/$\mu$s) back to
172~G,
efficiently converting the atoms to molecules. For
modest densities ($\langle\rho\rangle\lsimeq
10^{12}\,{\rm cm}^{-3}$),
ro-vibrational quenching at a rate comparable to
Na~\cite{MA_MBEC} should
be tolerable on millisecond
timescales~\cite{COLLAPSE}, and the efficient
production of a molecular BEC ought to be within easy
reach of
experiments.

We have come full circle. In our earliest
publication on coherent photoassociation (and, by
formal
equivalence, on magnetoassociation) of a BEC, we
predicted rapid
adiabatic passage from an atomic to a molecular
condensate
induced by a sweep of the detuning $\delta$ across the
resonance \cite{JAV99}. Those arguments were based on
a two-mode model, i.e., atomic and molecular
condensates only. In the interim it has become evident
\cite{KOS00,JAV02}
that rogue dissociation to noncondensate atoms is a
dominant factor in the
dynamics whenever $\Omega\gsimeq\omega_\rho$, a
condition met by two
orders of magnitude in
the JILA experiments \cite{COR00,DON01,CLA02,DON02}.
Hence, rapid adiabatic passage was not a leading
possibility for creating
a $^{85}\rm{Rb}_2$ molecular condensate. Now it has
made a comeback with a vengeance.

While preparing this manuscript, we became aware of
similar
Na work by Yurovsky and Ben-Reuven~\cite{YUR02}.

Support: Academy of
Finland (projects 43336 and 50314: MM and KAS);
NSF and NASA (PHY-0097974, and NAG8-1428: AC, MK,
CX, YZ, and JJ);  and NSF REU program (RJP);
express thanks (from JJ) to the Helsinki University of
Technology and
Matti Kaivola for support and hospitality during
completion of this work.

\begin{references}
\vspace{-0.5 cm}

\bibitem{UNIT_EFF}
   J. Javanainen and M. Mackie, \pra {\bf 58}, R789 (1998);
   P.S.Julienne \etal, \pra {\bf 58}, R797 (1999).

\bibitem{MA_MBEC}
E. Timmermans \etal, Phys. Rep. {\bf 315}, 199 (1999).

\bibitem{JAV99}
J. Javanainen and M. Mackie, \pra {\bf 59}, R3186
(1999).

\bibitem{RAMAN}
D. J. Heinzen \etal, \prl {\bf 84}, 5029 (2000).

\bibitem{KOS00}
M. Ko\u{s}trun \etal, \pra {\bf 62}, 063616 (2000).

\bibitem{JAV02}
J. Javanainen and M. Mackie, \prl {\bf 88}, 090403
(2002).

\bibitem{BOH99}
For similar collision-theory results, see
J. L. Bohn and P.~S. Julienne, \pra {\bf 60}, 414
(1999).

\bibitem{OTHER_ROGUES}
K. G\'oral \etal, \prl {\bf 86}, 1397 (2001);
M.~J. Holland \etal, {\it ibid.} {\bf 86}, 1915 (2001).

\bibitem{WYN00}
R. Wynar \etal, Science {\bf 287}, 1016 (2000).

\bibitem{MCK02}
C. McKenzie \etal, \prl {\bf 88}, 120403 (2002).

\bibitem{PIC02}
M. Pichler and R. Hulet, (private communication).

\bibitem{COR00} S. L. Cornish \etal, \prl {\bf 85},
1795 (2000).

\bibitem{DON01}
E. A. Donley \etal, Nature (London) {\bf 412}, 295
(2001).

\bibitem{CLA02}
N. R. Claussen \etal, \prl {\bf 89}, 010401 (2002).

\bibitem{DON02}
E. A. Donley \etal, Nature (London) {\bf 417},  529
(2002).

\bibitem{MAC02}
M. Mackie \etal, \prl {\bf 89}, 180403 (2002).

\bibitem{KOK02}
   S. J. J. M. F. Kokkelmans and M. J. Holland, \prl {\bf
89}, 180401 (2002).

\bibitem{KOE02}
T.~K\"ohler \etal, \pra {\bf 67}, 013601 (2003).

\bibitem{DUI02}
  R. A. Duine and H.~T.~C. Stoof, (cond-mat/0210544).

\bibitem{LZMA}
F. H. Mies \etal, \pra {\bf 61}, 022721 (2000);
V. A. Yurovsky \etal, \pra {\bf 62}, 043605 (2000).

\bibitem{COLLAPSE}
Collapse due to the switch from below to above
threshold, i.e., from positive to negative scattering
length, presumably occurs on the
trap-dynamics timescale ($\gsimeq 100\,{\rm
ms}$), and has no time to set in.

\bibitem{YUR02}
V.A. Yurovsky and A.Ben-Reuven, (cond-mat/0205267).

\end{references}
\end{document}